\documentclass[12pt]{article}
\pdfoutput=1

\usepackage{mathtools}
\usepackage{bm}
\usepackage{float}
\usepackage{relsize}
\usepackage{array}
\usepackage{fullpage}
\usepackage{graphicx}
\usepackage{psfrag}
\usepackage{epsf}
\usepackage{amsfonts}
\usepackage{amsmath}
\usepackage{pstricks}
\usepackage{color}
\usepackage{setspace}
\usepackage{easybmat}
\usepackage{slashed}
\usepackage{amsmath}
\usepackage{amssymb}
\usepackage{graphicx}
\usepackage{verbatim}
\usepackage{caption}
\usepackage{subcaption}
\usepackage{multirow}
\usepackage[square,sort,comma,numbers]{natbib}
\usepackage{tikz}
\usepackage{latexsym}
\usepackage{mathptmx}
\usepackage{lineno}
\usepackage{amsmath}
\usepackage{amsfonts}
\usepackage{amssymb}{}
\usepackage{amsbsy}
\usepackage{bm} 
\usepackage{latexsym}
\usepackage{graphicx}
\usepackage{mathptmx}
\usepackage{xcolor}
\usepackage[symbol]{footmisc}
\usepackage{eucal}
\usepackage{enumitem}
\usepackage{ragged2e}
\usepackage{changepage}
\usepackage[T1]{fontenc}                
\usepackage[utf8]{inputenc}  
\usepackage{lineno} 
\usepackage[english]{babel}
\usepackage{tikz}
  \usetikzlibrary{shapes,arrows,fit,calc,positioning}
  \tikzset{box/.style={draw, diamond, thick, text centered, minimum height=0.5cm, minimum width=1cm}}
  \tikzset{line/.style={draw, thick, -latex'}}
\usepackage{blkarray}
\usepackage[hypcap=false]{caption}
\usepackage{subcaption}
\usepackage[pdftex,colorlinks=true,urlcolor=blue,citecolor=black,anchorcolor=black,linkcolor=black,bookmarks=false]{hyperref}

\usepackage[numbers]{natbib}
\usepackage{authblk}

\usepackage{mathrsfs}
\DeclareMathAlphabet\mathbfcal{OMS}{cmsy}{b}{n}
\makeatletter
\newcommand*\bigcdot{\mathpalette\bigcdot@{.5}}
\newcommand*\bigcdot@[2]{\mathbin{\vcenter{\hbox{\scalebox{#2}{$\m@th#1\bullet$}}}}}
\makeatother

\usepackage[ruled, vlined, linesnumberedhidden, resetcount, english]{algorithm2e}

\newcommand{\ignore}[1]{}
\numberwithin{equation}{section}

\title{An Information Theory Treatment of Animal Movement Tracks$^\dag$}
\author[1,2]{Wayne M. Getz\footnote{Correspondence: \href{mailto:wgetz@berkeley.edu}{wgetz@berkeley.edu} \\ 
$^\dag$Prepared for the Springer volume ``The Mathematics of Movement: An Interdisciplinary Approach to Mutual Challenges in Animal Ecology and Cell Biology,'' Edited by Luca Giuggioli \& Philip Maini}}
\affil[1]{Dept.~Environmental Science, Policy and Management, University of California, Berkeley, 94720 CA, USA}
\affil[2]{School of Mathematics, Statistics \& Computer Science, University of KwaZulu-Natal, South Africa}

\date{}                     
\begin{document}
\fontfamily{pvh}\selectfont
\maketitle
\vspace{-1.3cm}
\begin{abstract}

Position recordings of the two-dimensional tracks of animals moving over landscapes has progressed over the past three decades from hourly to second-by-second locations.  Track segmentation methods for analyzing the behavioral information in such relocation data has lagged somewhat behind, with scales of analysis currently at the sub-hourly to minute level. A new approach is needed to bring segmentation analysis down to a second-by-second level. Here, a fine-scale approach is presented that rests heavily on concepts from Shannon's Information Theory. In this paper, we first briefly review and update concepts relating to movement path segmentation. We then discuss how cluster analysis can be used to organize the smallest viable statistical movement elements (StaMEs), which are $\mu$ steps long, and to code the next level of movement elements called ``words'' that are $m \mu$ steps long. Centroids of these word clusters are identified as canonical activity modes (CAMs).  Unlike current behavioral change point analysis and hidden Markov model segmentation schemes, the approach presented here allows us to provide entropy measures for movement paths, compute the coding efficiencies of derived StaMEs and CAMs, and to assess error rates in the allocation of strings of $m$ StaMEs to CAM types. In addition our approach allows us to employ the Jensen-Shannon divergence measure to assess and compare the best choices for the various parameters (number of steps in a StaME, number of StaME types, number of StaMEs in a word, number of CAM types), as well as the best clustering methods for generating segments that can then be used to interpret and predict sequences of higher order segments. The theory presented here provides another tool in our toolbox for dealing with the effects of global change on the movement and redistribution of animals across altered landscapes.\\[12pt]

\end{abstract}

\section{Introduction}

At its most fundamental level, the movement track $\mathcal{T}$ of an animal over a landscape is a time series of position relocation points. Movement track relocation data conceal a movement behavior narrative that, once revealed, provides insights into and a deeper knowledge of the ecological aspects of an individual's life history \citep{nathan2022big}. Extracting this knowledge requires, first, a rigorous decoding to reveal the movement elements underpinning the relocation data.  Second, it requires an understanding of how these movement elements are influenced by both the internal state (age, sex, health, physiology) of an individual and its external surroundings (landscape, climatic, and population ecological factors).  

These movement elements themselves cannot be identified by their mechanical signature (i.e., fundamental movement elements in the sense described in \cite{getz2008framework}) because such information is not coded in relocation data alone, but requires either video, accelerometer or other body-parts movement data for identification \citep{nathan2012using}.  Rather they can be \textcolor{black}{replaced by statistically derived movement elements computed from a fixed number of relocation data steps}, using a recent approach described in \citet{getz2023statistical}.  \textcolor{black}{This approach permits a coding theory of the relocation data to be formulated, as  developed here,  though the approach} requires relatively high frequency relocation data that has an inter-point interval of, at most, seconds rather than minutes (i.e. frequencies in excess of $0.1$ Hz).

Decoding a movement track to read the story it encodes involves a process called path segmentation \citep{edelhoff2016path}. This process relies on time series segmentation methods such as behavioral change point analysis (BCPA) \citep{gurarie2009novel,chen2011parametric,gurarie2016animal,Teimouri2018clustering,gundermann2023change,thompson2024simultaneous} and hidden Markov methods (HMM) \cite{franke2004analysis,langrock2012flexible,michelot2016movehmm,zucchini2016hidden,pohle2017selecting} \textcolor{black}{that produce variable length segments. Typically, such segments are much longer than the statistical elements discussed in \citet{getz2023statistical}; and we note that if a code is not based on fixed-length segments, then ambiguity arises regarding the number of elements used to code any given segment of track.} Additionally, BCPA and HMM methods, as mentioned by \citet{gundermann2023change}, may not be the best way to identify rare events, such as ``parturition, migration initiation, and juvenile dispersal.''

The decoding process is generally not applied directly to the relocation data, but first these data are transformed into a time series of step lengths (SL) and turning angles (TA) \cite{seidel2018ecological,getz2022hierarchical,thompson2024simultaneous}.  From these, other quantities can be extracted, such as the individual's persistent and angular velocities \cite{gurarie2009novel}. These quantities can then be used to identify intervals of time over which they vary in characteristic ways.  Through clustering or other category generating procedures \cite{kassambara2017practical,jaeger2023cluster}, segments with particular characteristic variations can be interpreted as expressing some identifiable behavioral movement mode, such as walking, resting, or feeding.

BCPA has proven to be a potent segmentation technique on sub-minute data. For example, as an illustration of its accuracy at this level of resolution, \citet{Teimouri2018clustering} used BCPA to segment relocation data collected every 10 seconds from four sheep grazing at an experimental site in Norway. To reduce global navigation satellite systems errors, though, they smoothed their data to obtain a relocation point every 1 min, then applied a 30 point (i.e. half hour) moving window to generate both persistent and angular velocities at each point.  This approach allowed them to identify four types of segments using Ward's agglomerative hierarchical clustering method \citep{murtagh2014ward}: ``foraging,'' ``resting,'' ``walking,'' and ``other.'' The segments they identified were all of variable length to a resolution of 1 minute. After ground-truthing a number of their segments, \citet{Teimouri2018clustering} found there method scored an average classification accuracy of around 80\% across their movement modes. 

For very large, high-resolution relocation data sets, BCPA may be relatively computationally expensive and also have problems converging on a solution, particularly if one seeks to identify four or more movement behavioral modes. An alternative approach based on Shannon coding theory can be taken. This approach is computationally much more manageable than BCPA and also provides a rigorous theory for measuring and comparing the information content (entropy) of the movement track segmentation process.

In 1948, Claude Shannon presented his mathematical theory of communication in a two part article published in the Bell System Technical Journal \citep{shannon1948mathematicala,shannon1948mathematicalb}. This theory can be applied to any sequential strings of symbols used to code information, whether they be strings of electronic bits, nucleic acids, numbers and letters, or animal movement track segments. In the latter case, however, the segments underlying the analysis (i.e., the lowest set of coding symbols) need to be standardized to all be of the same size to avoid confusion between single and compound segments of relocation data. 

In the material that follows, we review an approach to segmenting  movement tracks using elements that are $\mu$ steps long and, after clustering and identifying cluster centroids, provides a set of $n$ statistical movement elements (StaME set ${\mathcal S}$, elements $\sigma_i$, $i=1,\cdots,n$). The approach also segments  movement tracks using elements referred to as ``words'' that are $m \mu$ steps long. After clustering these words, there centroids, provide a set of $k$ canonical activity modes (CAM set ${\mathcal K}$, elements $\kappa_c$, $c=1,\cdots,k$). The smaller StaMEs may then serve as set of $m$ symbols for analyzing the information content of the larger CAMs that, in turn, can be used to probe the information content of subdiel, diel, and supra-diel segments of animal movement tracks (Table 1). 

Our formulation also allows us to compare coding efficiencies across different methods for clustering and generating StaMEs and CAMs. This includes the effects of the number of steps $\mu$ selected for the StaME building blocks of movement tracks, the number $n$ of symbol types identified, the number $m$ of smallest segments used to build words, and the number $k$ of CAMs used to generate larger interpretable movement behavior segments, such as behavioral activity modes and diel activity routines \cite{getz2022hierarchical,luisa2023categorizing}.

\section{Segmentation Hierarchy}

\begin{table}[h]
\caption{Hierarchical segmentation of an animal track relocation time series $\mathcal{T}^{\rm loc}$ collected at a frequency $f$ HZ; names and resolutions provided here are a more precise rendering of previous presentations \cite{getz2022hierarchical,getz2023animal} (see footnotes).}\label{tab:SCBD}
    \begin{tabular}{|l|cll|} \hline
    &&& \\[-6pt]
    {\bf Acronym} &  {\bf Number of} & \raisebox{-6pt}{\bf Resolution} & \hspace{1.5cm} \raisebox{-6pt}{\bf Comment} \\[-4pt] 
   (symbol) & {\bf types} && \\[4pt] \hline
     &&& \\[-6pt]
      StaME$^*$ $(\sigma_i)$  & $i=1,\cdots,n$  & $\mu f$ secs & $n$ clusters of segments and $\mu$ steps   \\ 
      & & & \raisebox{4pt}{per segment are method parameters} \\[-2pt]
       CAM$^{\dag}$ $(\kappa_c)$  & $c=1,\cdots,k$ &  $m \mu f$ secs & $k$ clusters of words and  $m$ StaMEs \\ 
       & & & \raisebox{4pt}{per word are method parameters} \\[-2pt]
        BAM$^{\ddag}$  & TBD$^{**}$  & sub-diel  &  homogeneous: string of same CAMs \\
         &   & \raisebox{4pt}{but variable}  & \raisebox{4pt}{heterogeneous: characteristic CAM mix} \\[-2pt]
       DAR$^\S$   & TBD$^{\dag\dag}$  & 24-hrs & bottom up: strings of BAMs \\
        &   & \raisebox{4pt}{fixed}  & \raisebox{4pt}{top down: e.g., geometry of DARs }\\ 
        LiMP$^\P$  & depends on  & multi-day to & LiMP specific DARs condition on \\
        &  \raisebox{4pt}{age \& env.$^{\ddag\ddag}$} & \raisebox{4pt}{seasonal} & \raisebox{4pt}{internal and external factors} \\ 
         LiT$^\parallel$  & syndromic  & individual's & LiMP specific DARs condition on \\
        & \raisebox{4pt}{move. types$^{\S\S}$}  & \raisebox{4pt}{life span} & \raisebox{4pt}{internal and external factors}\\
        \hline
    \end{tabular}
    \vspace{6pt}
    
{\footnotesize 
$^*$Statistical movement elements, formerly called metaFuMEs \\[-3pt]
$^{\dag}$Canonical activity mode segments, formerly called ``short duration'' or homogeneous CAMs  \\[-3pt]
$^{\ddag}$Behavioral activity mode segments, formerly called ``long duration'' or heterogeneous CAMs \\[-3pt]
$^\S$Diel activity routine segments \\[-3pt]
$^\P$Life-time movement phase segments, including dispersal, migration, and nurturing young\\[-3pt]
$^\parallel$Life-time track, including dispersal, migration, and nurturing young\\[-3pt]
$^{**}$To be determined using biological change point analysis \citep{gurarie2009novel,chen2011parametric} and hidden Markov methods \cite{franke2004analysis,michelot2016movehmm,zucchini2016hidden} \\[-3pt]
$^{\dag\dag}$May be extracted using appropriate clustering methods \cite{luisa2023categorizing} \\[-3pt]
$^{\ddag\ddag}$To be determined using appropriate statistical methods \citep{owen2014coping,klarevas2024diel} \\[-3pt]
$^{\S \S}$For discussions on movement syndrome types see \citep{abrahms2017suite,spiegel2017}}
\end{table}

A hierarchical scheme for the segmentation of tracks has been discussed in considerable depth elsewhere \citep{getz2022hierarchical,getz2023animal,getz2023statistical}. An updated summary is provided in Table 1. Originally, \citet{getz2008framework} proposed that, at its most basic level, a movement track $\mathcal{T}$ can be conceptualized as a sequence of fundamental movement elements (FuMEs; referred to as FMEs in \cite{getz2008framework}). These elements when strung together either homogeneously (or one type), or in characteristic mixes of several types, produce canonical activity mode (CAMs) segments.  \citet{getz2008framework} pointed out that ``with most current data, it is not possible to construct distributions of building block movement elements in terms of FuMEs'' (they used the acronym FME in \cite{getz2008framework}):  essentially, as previously mentioned, FuMEs are unobservable in relocation data alone and companion video or accelerometer data are likely needed to identify the start of each new FuME. This subsequently led to the concept of a set $\mathcal{S}$ of statistical movement elements (StaMEs), formerly referred to as a metaFuMEs \cite{getz2022hierarchical} to be used in place of FuMEs as a set of lowest level movement building block elements or coding symbols.   In the formulation here, the set $\mathcal{S}$ is regarded as an underlying set of symbols that can be used to encode the movement track relocation data time series $\mathcal{T}^{\rm loc}$ as a StaMEs time series $\mathcal{T}^{\sigma}$ (Table 2).  This symbolic representation then allows an information theory toolbox to be used to analyze the information content and, ultimately, meaning of animal movement tracks \cite{mayner2018pyphi}. 

\begin{figure}[t!]
    \centering
    \includegraphics[width=1.0\textwidth]{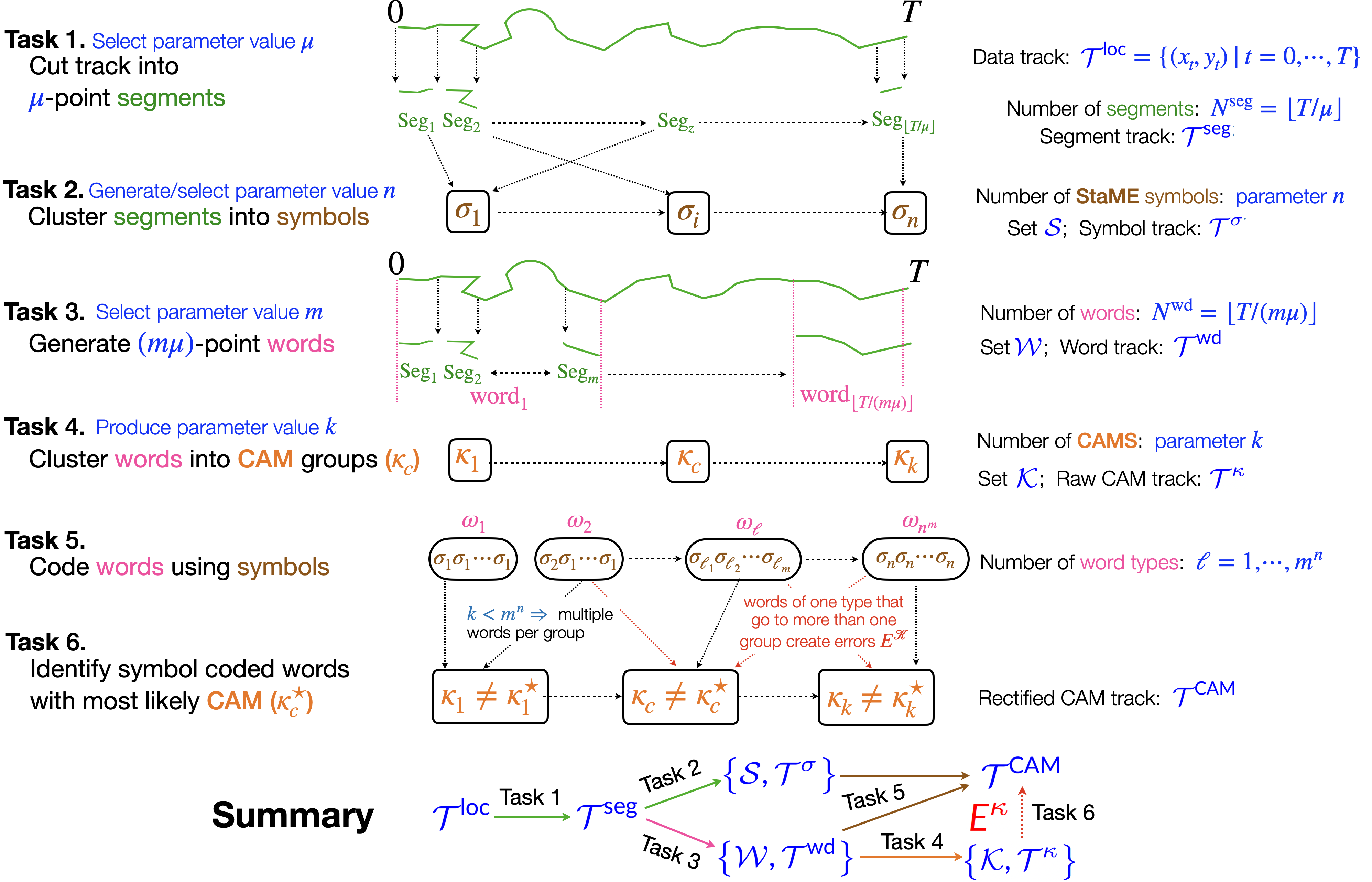}
    \caption{\small {\bf A.} A graphic depiction of tasks 1-6 required to code an animal movement relocation data series ($\mathcal{T}^{\rm loc}$) into a  statistical movement elements (StaMEs) data series $\mathcal{T}^{\sigma}$ and a canonical activity modes (CAMs) data series $\mathcal{T}^{\rm CAM}$ and compute the coding accuracy $E^{\mathcal{\kappa}}$ of the process: i.e., implement the functor $\mathcal{M}:\mathcal{T}^{\rm loc} \mapsto \left\{\mathcal{T}^{\sigma},\mathcal{T}^{\rm CAM},E^{\mathcal{\kappa}} \right\}$.  Parameters values $\mu,n,m,$ and $k$ need to be either {\it a priori} selected or determined during implementation of the clustering methods are highlighted in blue.  Extracted objects are: segments in green (Tasks 1 and 3), symbols in brown (Task 2), words in fuchsia (Tasks 3-5), CAMs in orange (Tasks 4 and 6), and a coding accuracy measure in red (Task 6).  With regard to the latter, the red dotted arrows between Tasks 5 and 6 indicate misassignment of a proportion of the $N_{\ell}$ words of type $\omega_\ell$ ($\ell=1,\cdots,n^m$; $N^{\rm wd} = \sum_{\ell=1}^{n^m} N_\ell$) as instances of particular CAMs assigned to $\kappa_{c_1}^\star$ when initially a member of $\mathcal{W}_{c_2}$ for $c_1 \ne c_2$, $c_1,c_2 \in \{1,\cdots,k\}$). A {\bf summary} of the tasks and objects produced at each step of the process is provided at the bottom of this graphic. (Note:  calligraphy letters in the graph and caption are the same symbol, but generated by different font sets)}
    \label{fig:Segmentation}
\end{figure}

The elements of $\mathcal{S}$, as discussed in some detail in \citet{getz2023statistical}, are generated from a segmentation of the $T+1$ point (0 included) relocation time series $\mathcal{T}^{\rm loc}$ into a track segment time series $\mathcal{T}^{\rm seg}$ consisting of $\lfloor T/\mu \rfloor$ ($\lfloor \bigcdot \rfloor$ means round down the integer below) segments each containing $\mu$ steps (Task 1, Fig~\ref{fig:Segmentation}). A cluster analysis of these segments is then used to place them into one of $n$ similarly-shaped categories of segment types  \citep{getz2023statistical}. The centroids of these $n$ segment-type clusters can then be extracted and treated as archetypal elements in a set $\mathcal{S}$ of symbolic segments $\sigma_i$ that can be used to code a time series $\mathcal{T}^{\sigma}$ (Task 2, Fig~\ref{fig:Segmentation}). The values $\mu$ and $n$ themselves become tune-able parameters of the StaME/symbol identification and creation process, while the method used to cluster the segments into a set of $n$ StaMEs becomes one of the arguments of the segmentation process when viewed as a functor $\mathcal{M}$ acting on the relocation data time series $\mathcal{T}^{\rm loc}$ (Table 2; also see Eq~\ref{eq:Method} below).

\begin{table}[!t]
\caption{Representation of movement tracks at different stages and resolutions of encoding  the relocation data $\mathcal{T}^{\rm loc}$ into StaMEs/symbols and CAMs (Fig~\ref{fig:Segmentation})}\label{tab:tracks}
    \begin{tabular}{|l|lll|} \hline
    &&& \\[-6pt]
    {\bf Track} (description) & {\bf Elements}   & {\bf Relevant Measures} & {\bf Comment} \\[4pt] \hline
     &&& \\[-6pt]
      $\mathcal{T}^{\rm loc}$ (point location track) & $\big(t;(x_t,y_t)\big)$ & $T+1$ points  & relocation data   \\[4pt] 
    $\mathcal{T}^{\rm seg}$ (segment track) & $(z;{\rm seg}_z)$  & $\mu$-steps; $\lfloor \frac{T}{\mu} \rfloor$ segments & 1st segmentation   \\[4pt] 
    $\mathcal{T}^{\sigma}$ (StaME track)& $(i;\sigma_i)$ & $\lfloor \frac{T}{\mu} \rfloor  H(\mathbf{p}^\sigma)$ bits$^*$  & 1st level centroids   \\[4pt]
    $\mathcal{T}^{\rm wd}$ (word track) & $(j;{\rm wd}_j)$  & $m \mu$-steps;  $\lfloor \frac{T}{m\mu} \rfloor$ words & 2nd segmentation   \\[8pt]
     $\mathcal{T}^{\kappa}$ (raw CAM track) & $(j;\kappa_{c})$  & $D_{\rm JS}^{\rm ens}\left({\mathbf p}^{\mathcal{W}_1},\cdots,{\mathbf p}^{\mathcal{W}_k}\right)^{\dag}$ & centroids of ${\rm wd}_j \in \mathcal{W}_c$   \\[8pt]
      $\mathcal{T}^{\rm CAM}$ (rectified CAM track) & $(j;\kappa_{c}^\star)^{+}$  & $\lfloor \frac{T}{m\mu} \rfloor  H(\mathbf{p}^{\rm CAM})$ bits$ ^\S $ & argmax $c$ reassignment   \\[4pt] \hline
    \end{tabular} 
\vspace{6pt}

{\small $^+$The difference between raw ($\kappa_c$) and rectified ($\kappa_c^\star$) CAMs based on Eq~\ref{eq:camAssign} (see Fig~\ref{fig:Segmentation})\\
$^*$Information encoded into $\mathcal{T}^{\rm loc}$ if symbol sequences are uncorrelated \\
$^{\dag}$Divergence across the ensemble distributions $\mathbf{p}^{\mathcal{W}_c}$, $c=1,\cdots,k$ (Eq~\ref{eq:JSDavg})\\
$^\ddag$Avg.~divergence of $\omega_\ell$ distributions in $\mathcal{W}_c$ from their distribution in $\mathcal{W}$ itself \\
$^\S$Information that can be coded into $\mathcal{T}^{\rm loc}$ if CAM sequences are uncorrelated}
\end{table}

The current concept of a CAM also needs to be treated more rigorously than before if it is to be integrated into an information theoretic framework. Specifically, in the formulation below, $m$ $\mu$-length segments are strung together into $m\mu$-step words (Task 3; Fig~\ref{fig:Segmentation}, Tables 1 and 2) that then constitute a word set $\mathcal{W}$. The words in $\mathcal{W}$ can then be used to generate a word time series $\mathcal{T}^{\rm wd}$ of $\lfloor T/(m\mu) \rfloor$ elements (Table 2). A cluster analysis of the words in $\mathcal{W}$ is then used to identify $k$ word clusters denoted by $\mathcal{W}_c$, $c=1,\ldots,k.$ The centroids of these $k$ clusters can be used to generate the $k$-element raw CAM set $\mathcal{K}$ (Task 4, Fig~\ref{fig:Segmentation}), with elements $\kappa_c$ 
used to code track $\mathcal{T}^{\rm wd}$ into a raw CAM track $\mathcal{T}^\kappa$ (Table 2).

Additionally, one may also translate each word into its representative $m$-symbol string and then identify which cluster most words that have the same $m$-symbol code belong to.  If this is the word set $\mathcal{W}_c$, then all words with this particular $m$-symbol coding sequence will be identified with a rectified CAM type $c$, as represented by its archetype/ideal $\kappa^\star_c$ (Task 5, Fig~\ref{fig:Segmentation}). The corresponding rectified CAM coding track is denoted by $\mathcal{T}^{\rm CAM}$.  The proportion $E^{\mathcal{\kappa}}$ of these particular $m$-symbol words that had to be reassigned, when added for all $c=1\cdots,k$, provides a measure of the reliability of the CAM coding process. In a nutshell, the difference between the  $\mathcal{T}^\kappa$ and $\mathcal{T}^{\rm CAM}$ time series is that the former arises from a cluster analysis of segments containing $m \mu$ steps and the latter assigns a CAM type $\kappa^\star_c$ to all words that are coded by the same $m$-symbol sequence of StaMEs $\sigma_i \in \mathcal{S}$.  This distinction will be made more rigorous in the next section where we formalize the set of Tasks 1-6 (Fig~\ref{fig:Segmentation}) that defines the functor $\mathcal{M}:\mathcal{T}^{\rm loc} \mapsto \left\{\mathcal{T}^{\sigma},\mathcal{T}^{\rm CAM},E^{\mathcal{\kappa}} \right\}$ used to derive the StaME and rectified CAM sets and track codings, and the assignment error associated with the process.

The frequencies of StaMEs and CAMs respectively coded into the movement track representations $\mathcal{T}^{\sigma}$ and $\mathcal{T}^{\rm CAM}$ can be used to derive a measure of the entropy of these coding schemes if coding elements are not autocorrelated.  If they are, then an autocorrelation analysis can be used to derive more accurate measures of the entropy of these coded tracks. Further, if CAMs are regarded as each consisting of a particular sequence of $m$ StaMEs, then it becomes  relatively simple to apply Shannon's Information Theory \cite{shannon1948mathematicala,shannon1948mathematicalb} concepts to compare the information content per unit time of different tracks.  

Although, each of the CAMs will come to be identified in terms of some emblematic activity such as resting, random stepping, directed stepping, jogging/cantering, sprinting/galloping, individual CAMs will not constitute a complete behavioral repertoire of a particular kind. Rather, it will take either a homogeneous or a characteristic-heterogeneous-mix of CAM strings of, in practice, variable length to then interpret the resulting multi-CAM segment as a particular kind of behavioral mode \cite{getz2022hierarchical,getz2023animal}.  For example, in the context of a homogeneous CAM string, a sequence of `high velocity, low turning angle' CAMs may be interpreted as a `beelining' behavioral activity mode (BAM).   In the context of a characteristic mix of CAMs, a mix of stepping and resting CAMs may come to be ground-truthed as a foraging BAM. Since BAMs are generally of variable length, they currently constitute the types of movement mode structures identified using BCPA methods \citep{gurarie2009novel,chen2011parametric,gurarie2016animal,Teimouri2018clustering,gundermann2023change,thompson2024simultaneous}.

From a top down viewpoint, BAMs are variable length behavioral segments that switch intermittently during the course of animal's diurnal activity routine (DAR). The lengths and frequency of switching among BAMs depends on many factors including, as pointed out in our introduction, the internal state of an individual (e.g., its age, sex, health, physiology) and its external environment (e.g., landscape and vegetation structure of surroundings, weather variables, and presence of particular conspecifics, heterospecific competitors, and predators).  The sequencing and switching probabilities of BAMs can be modeled by fitting an autoregressive time series (AR)  CAM occurrence model to empirical movement path data ${\mathcal T}^{\rm loc}$ that has been segmented into a rectified CAM times series $\mathcal{T}^{\rm CAM}$.  Such models can also be elaborated by including the effects of internal and external auxiliary variables on sequencing and switching rates.  

Methods for fitting categorical variable AR($p$) models to empirical data (here $p$ denotes the time span of the correlated effect---but to begin it may be sufficient to fit an AR(1) model). Dependence of model parameters on auxiliary factors is likely to be important and models that deal with such situations can be used (e.g., \citep{biswas2009time}). Model parameters influencing CAM switch probabilities in generalized AR$(p)$ models are likely to be time dependent, particularly with regard to BAMs that occur at characteristic times of the day or night within the diel cycle (e.g., resting, heading to water, heading home, and feeding).  Seasonal effects are also likely to be a factor and can be treated under the rubric of segmenting the lifetime track  (LiT) of individuals into lifetime movement phases (LiMPs), as they may pertain, for example, to breeding, hibernation, and migration cycles.

\section{Two Level Track Segmentation}
An animal movement track $\mathcal{T}$ can viewed from a number different perspectives.  In a two-level hierarchical segmentation of the time series $\mathcal{T}^{\rm loc}$ of relocation points $(t;x_t,y_t)$, $t=0,\cdots,T$, derived time series, as previously mentioned, can be viewed as (Fig~\ref{fig:Segmentation}, Table 2): i) a time series ${\mathcal T}^{\rm seg}$ of building-block segments, where each segment consists of $\mu$ consecutive steps; ii) a  time series $\mathcal{T}^{\sigma}$ of segments labeled by their StaME (symbol) type, as it arises from a clustering of segments in ${\mathcal T}^{\rm seg}$ into $n$ distinct categories; iii) a time series ${\mathcal T}^{\rm wd}$ of word segments, where each segment consists of $m \mu$ consecutive steps; iv)  a time series ${\mathcal T}^{\kappa}$ of raw CAMs that arise from clustering $(m \mu)$-step length words into $k$ CAM categories; and v)  a time series  ${\mathcal T}^{\rm CAM}$ of rectified CAMs that arises from reassigning words, based on their $m$-length string of StaME coding symbols $\sigma_i$ to the CAM where such words are most likely to occur in the first place.  Once this segmentation has been accomplished then different methods can be used to parse the rectified CAM time series ${\mathcal T}^{\rm CAM}$ into variable length BAMs, fixed time periods DARS, and multi-day lifetime movement phases (LiMPs) using autoregressive modeling, BCPA and HMM methods. These various coding schemes for $\mathcal{T}$ then allow one to use Shannon information theory to estimate coding entropy and error rates and compare the coding accuracy of different coding schemes.

In summary, the six tasks needed to implement the functor $\mathcal{M}:\mathcal{T}^{\rm loc} \mapsto \left\{\mathcal{T}^{\sigma},\mathcal{T}^{\rm CAM},E^{\mathcal{\kappa}} \right\}$ are (Fig~\ref{fig:Segmentation}):
\begin{itemize}
    \item[] {\bf Task 1.} \underline{Movement Path Segmentation:} 
    This task requires that we segment the movement track relocation data set
    \begin{equation}\label{eq:trackloc}
        {\mathcal T}^{\rm loc}=\{(t;x_t,y_t) | t=0,\cdots,T\}
    \end{equation}
     into segments of size $\mu$ (i.e. each consists of $\mu$ steps) to obtain a movement track representation ${\mathcal T}^{\rm seg}$ of $N^{\rm seg}=\lfloor T/\mu \rfloor$ segments ($\lfloor {\bigcdot} \rfloor$ means round down to the integer value): i.e., 
     \begin{equation}\label{eq:trackseg}
         {\mathcal T}^{\rm seg}  =  \big\{(z;{\rm seg_{z}}) | {\rm seg}_{z} = \big((x_{\mu (z-1)},y_{\mu (z-1)}),\cdots,(x_{\mu z },y_{\mu z})\big), \ z=1,\cdots,N^{\rm seg} \big\} 
    \end{equation}

    \item[] {\bf Task 2.} \underline{Segment Clustering:} 
    
    This task produces a set StaMEs
    \begin{equation}\label{eq:StameSet}
        {\mathcal S}=\{\sigma_i | i=1,\cdots,n \} 
    \end{equation}
    that can be used as a set of symbols to recode the set of $T+1$ relocation points $\mathcal{T}^{\rm loc}$ as a times series $\mathcal{T}^{\sigma}$ of $\lfloor T/\mu \rfloor$ symbols (Table 2).
    The process of creating the set ${\mathcal S}$ requires that we cluster the segments making up the track ${\mathcal T}^{\rm seg}$ using either shape (typically, unsupervised machine learning \cite{alloghani2020systematic}) methods or vector methods that use summary statistics as variables defining each segment \citep{kassambara2017practical,sethi2024information}.  Shape methods my be set up to ignore rotations and reflections when movement on a landscape is unlikely to be directionally and rotationally biased. Biases enter when individuals have a proclivity to move latitudionally or longitudinally at a local scale (e.g. in neotropical birds \cite{jahn2020bird}), or clockwise or anticlockwise \cite{jacobs2010sense,narazaki2021similar}.  Vector methods in the past, using summary statistics, have included the statistical means of persistence and angular velocities \citep{gurarie2009novel,Teimouri2018clustering}, the means and standard deviations of the speed (i.e. step length) and absolute turning angle of each segment, and sometimes even the net displacement (distance between the two end points) of each segment \citep{getz2023statistical,sethi2024information}.  
    
    Once the set of symbols $S$ has been constructed---where, for convenience, symbols are numbered by size (i.e., by step-length/velocity averaged across their $\mu$ consecutive steps) so the ${\rm size}(\sigma_i) \ge {\rm size}(\sigma_{i+1})$, $i=1,\cdots,n-1$---then the segmented data series ${\mathcal T}^{\rm seg}$ can be coded using these symbols to obtain the symbol time series
     \begin{equation}\label{eq:tracksym}
        {\mathcal T}^{\sigma}=\{(z;\sigma_{i_z}) | z=1,\cdots,N^{\rm seg}\} 
    \end{equation}
    We can now count the number of times that each symbol $\sigma_i$ appears in $\mathcal{T}^{\sigma}$ to obtain the vector 
    \begin{equation}\label{eq:propsym}
        \mathbf{p}^{\sigma}=\left(p^{\sigma}_1\cdots,p^{\omega}_{n} \right)'\, , \ \ \mbox{where} \ \ p^\sigma_i  \mbox{ is the proportion of $\sigma_i \in \mathcal{T}^{\sigma}, i=1,\cdots,n$ }
    \end{equation}
   of the  distribution of symbols  in $\mathcal{T}^{\sigma}.$
    
    \item[] {\bf Task 3.} \underline{Word Generation:} 
    
    In the same way that ${\mathcal T}^{\rm loc}$ can be parsed into $N^{\rm seg}$ segments each consisting of $\mu$ steps, we can parse ${\mathcal T}^{\rm loc}$ into $N^{\rm wd}$ words each consisting of $m$ consecutive segments providing us with a next level of segmentation in which segments contain $m \mu$ consecutive steps.  In this case we obtain a time series of $N^{\rm wd}= \lfloor T/(m\mu) \rfloor$  words
     \begin{equation}\label{eq:trackword}
        \mathcal{T}^{\rm wd} = \big\{(j;{\rm wd}_j |
          {\rm wd}_j  =  \big((x_{ m\mu (j-1)},y_{m\mu (j-1)}\big),\cdots,  
          \big(x_{m \mu j},y_{m\mu j})\big),\ j=1,\cdots,N^{\rm wd}  \big\}
    \end{equation}
After this segmentation is complete, we can gather these segments into a set of words
    \begin{equation}\label{eq:setOfwords}
        \mathcal{W}=\big\{{\rm wd}_j \in {\mathcal T}^{\rm wd} | j=1,\cdots,N^{\rm wd} \big\}
    \end{equation}

 \item[] {\bf Task 4.} \underline{Direct Word Clustering and CAM Generation:}
    
 The set $\mathcal{W}$ of the words in the time series ${\mathcal T}^{\rm wd}$ can also be clustered to obtain a set of objects that we call raw canonical activity modes (CAMs).  Thus after clustering the $N^{\rm wd}$ words in $\mathcal{T}^{\rm wd}$ into $k$ different categories, using some appropriate vector or shape method,  we obtain $k$ subsets ${\mathcal W}_c \subset \mathcal{W}$, $c=1,\cdots,k$, each with $N^{\rm wd}_c$ words, where $\sum_{c=1}^k N^{\rm wd}_c = N^{\rm wd}$.  We can either regard each word segment $\omega_\ell$ ($\ell=1,\cdots,m^n$) in ${\mathcal W}_c$ as an instance of the  CAM $\kappa_c$ in the set
    \begin{equation}\label{eq:setOfCAMs}
    {\mathcal K}=\{\kappa_c \, | \, c=1,\cdots,k \},  \ \mbox{ and } \ \mathbf{p}^{\kappa}=\left(p^{\kappa}_{1},\cdots,p^{\kappa}_{k} \right)', \ \ \mbox{ where } \ \ p^{\kappa}_{c}=\frac{N^{\rm wd}_c}{N^{\rm wd}}
    \end{equation}
and $\kappa_c$ is the centroid  of all $\omega_\ell \in \mathcal{W}_c$.

Once all the words $\omega_\ell$, $\ell=1,\cdots,m^n$, have been assigned to some particular raw CAM group represented by the centroid $\kappa_c$ of the word cluster $\mathcal{W}_c$, then our movement track $\mathcal{T}^{\rm wd}$ can be rewritten as the raw CAM time series
\begin{equation}\label{eq:trackcam}
        {\mathcal T}^{\kappa}=\big\{(j;\kappa_{c}) | \mbox{ whenever } {\rm wd}_j \in \mathcal{W}_c \big\}
    \end{equation}
    
   \item[]  {\bf Task 5.}  \underline{Symbolic Word Coding:} 
   
   By comparing the way segments in the two time series ${\mathcal T}^{\sigma}$ and ${\mathcal T}^{\rm wd}$ have been labeled, we can time match each of the words ${\rm wd}_j \in {\mathcal T}^{\rm wd}$ with its corresponding coding sequence to obtain its representation
    \begin{equation}\label{eq:symbolWordEquiv}
        {\rm wd}_{j} \equiv \sigma_{j_1}\sigma_{j_2}\cdots\sigma_{j_m}, \ j=1,\cdots,N^{\rm wd}
    \end{equation}
    The number of different types of words that exist when $m$ symbols are strung together to code a word is $m^n$.  Thus we can generate the set ${\Omega}$ of all possible words that can be constructed from the set of $m$ symbols $\sigma_r \in {\mathcal S}$ (Eq~\ref{eq:StameSet}). The number of words in ${\Omega}$ is thus $n^m$ and 
    \begin{equation}\label{eq:wordSym}
       \Omega = \{\omega_\ell | \ell=1,\cdots,n^m\}, \ \ \mbox{where} \ \ \omega_{\ell} \equiv \sigma_{\ell_1}\sigma_{\ell_2}\cdots\sigma_{\ell_m} \ \ \mbox{and $\sigma_{\ell_r} \in {\mathcal S}$ for all $r=1,\cdots,m$}
    \end{equation}
    For convenience, the words in $\Omega$ are partially ordered so that the largest word is $\omega_1 \equiv \sigma_1 \cdots \sigma_1$, the smallest $\omega_{m^n} \equiv \sigma_n \cdots \sigma_n$  with the rest of the words going approximately from largest to smallest (in terms of the average step size $\equiv$ velocity taken across their consecutive steps) using an ordered numbering scheme, as discussed in \cite{sethi2024information}.  Once all these words have been numbered, we can  generate the proportions
    \begin{equation}\label{eq:propword}
        \mathbf{p}^{\omega}=\left(p^{\omega}_1\cdots,p^{\omega}_{m^n} \right)' \ \ \mbox{where} \ \ p^\omega_\ell  \mbox{ is the proportion of $\omega_\ell \in \mathcal{W}, \ell=1,\cdots,m^n$ }
    \end{equation}

\item[] {\bf Task 6.} \underline{Assign Coded Words to Rectified CAM:}

Beyond generating ${\mathcal T}^{\kappa}$ directly from  ${\mathcal T}^{\rm loc}$, we can identify different word types $\omega_\ell$ with specific CAMs $\kappa_c$ using some appropriate rule. For example, if we define $\mathbf{p}^{\mathcal{W}_c}$ as the vector of the proportions of $\omega_\ell$ words in the word clusters $\mathcal{W}_c$: i.e., 
\begin{equation}\label{eq:propwordC}
        \mathbf{p}^{\mathcal{W}_c}=\left(p^{\mathcal{W}_c}_1\cdots,p^{\mathcal{W}_c}_{m^n} \right)', \ 
        \  c=1,\cdots,k     
\end{equation}
then we can apply a rule, such as the following, to assign the various word types $\omega_\ell$ to particular rectified CAM types $\kappa_c^\star$
\begin{equation}\label{eq:camAssign}
  {\omega}_{\ell} \in \kappa_c^\star  \ \mbox{ whenever } c=\underset{c=1,\cdots,k}{\rm argmax} \, \{p^{{\mathcal W}_c}_\ell \}, \quad \ell=1,\cdots,m^n
\end{equation}
The star is used to denote the fact that the words $\omega_\ell$  that are identified with $\kappa_c^\star$ are different from the words $\omega_\ell \in \mathcal{W}_c$, whose centroid is our archetype/ideal $\kappa_c \in \mathcal{K}$ (Eq~\ref{eq:setOfCAMs}). In short, the archetypes $\kappa_c$ arise from a cluster analysis of a segmentation of $\mathcal{T}^{\rm loc}$ using segments of size $m \mu$.  On the other hand, the archetypes $\kappa_c^\star$ arise from a reassortment of the clusters $\mathcal{W}_c$ using rule Eq~\ref{eq:camAssign}. 
If there is more than one maximum value when applying rule Eq~\ref{eq:camAssign}, then we can select among them at random or using some additional cluster-related criterion.  

Once a rule such as Eq~\ref{eq:camAssign} has been applied to associate the $\ell$-th word position in ${\mathcal T}^\kappa$ with a particular $\kappa_c^\star$, then our time series and proportional occurrence can be recast as 
\begin{eqnarray}\label{eq:trackcamassign}
        \mathcal{T}^{\rm CAM} & = & \big\{(j;\kappa_{c}^\star) | \mbox{ whenever wd$_j$ is assigned to $\kappa_c$ using Eq~\ref{eq:camAssign} } \big\} \\
   && \mbox{with} \ \ {\mathcal K}^\star  =  \{\kappa^\star_c \, | \, c=1,\cdots,k \}   \ \ \mbox{and} \ \  \mathbf{p}^{\kappa^\star} = \left(p^{\kappa^\star}_1\cdots,p^{\kappa^\star}_{k} \right) \nonumber    
\end{eqnarray}

In general, we cannot expect $\mathcal{T}^{\kappa}$ and $\mathcal{T}^{\rm CAM}$ to represent the same sequence of CAMs.  Errors will occur and the proportion of errors that occur can be determined by computing the proportion of locations $j$ where the CAM designations in $\mathcal{T}^{\kappa}$ and $\mathcal{T}^{\rm CAM}$ do not match up.  This error computation can be either be computed by directly comparing the CAM sequences in $\mathcal{T}^{\kappa}$ and $\mathcal{T}^{\rm CAM}$ or using the  proportion of misclassifications that arise in computing the CAM assignment using rule Eq~\ref{eq:camAssign}.  In the latter case it is clear that the proportion of times that word$_j$, when coded as word type $\omega_\ell$, is correctly assigned to CAM $\kappa_c$ is $\left(\max\limits_{c} \{p^{{\mathcal W}_{c}}_j\}/\sum_{c=1}^{k} p^{{\mathcal W}_{c}}_j\right)$: proportional misassigned to another CAM is 1 minus this quantity.  Thus, summing over all misassignments of $\omega_\ell$, taking into account the  frequency $N^{\omega_\ell}/N^{\rm wd}$ of these words in $\mathcal{W}$ we obtain the percentage error rate 
\begin{equation}\label{eq:totError}
 E^{\kappa} = 100 \sum_{\ell=1}^{n^m}\frac{N^{\omega_\ell}}{N^{\rm wd}}\left(1-\frac{\max\limits_{c} \{p^{{\mathcal W}_{c}}_\ell\}}{\sum_{c=1}^{k} p^{{\mathcal W}_{c}}_\ell} \right) \%
\end{equation}
\end{itemize}

\section{Information Entropy of Movement Track Segments}

Information theory, as developed by Claude Shannon \cite{shannon1948mathematicala,shannon1948mathematicalb} in the late 1940's, posits a set of $n$ symbols ${\mathcal S}$ (Eq~\ref{eq:StameSet}) that when strung together sequentially encodes information at a basic rate determined by the probability $p^\sigma_i$ of encountering symbol $\sigma_i$ at randomly selected locations along an encoded string. 
Under these circumstances the information coding rate of the symbols in $\mathcal{S}$ is determined by its $\log_2$ entropy  $H\left({\mathbf p}^{\sigma}\right)$ \cite{manning1999foundations}
\begin{equation}\label{eq:infoSigma}
H^{\sigma} \equiv H\left({\mathbf p}^{\sigma}\right) = - \sum_{i=1}^n p_i^{\sigma} \log_2 p_i^{\sigma} \quad ({\rm bits})
\end{equation}
The units of entropy, when caste in the information theoretic  $\log_2$ context are ``bits'' of information with a single bit the rate at which information can be coded just using uncorrelated sequences of 0's and 1's with equal frequency (i.e., $-2 \times \frac{1}{2}\log_2\left(\frac{1}{2} \right) =1  $ bit).

Similarly, for the distribution ${\mathbf p}^{\omega}$ defined in Eq~\ref{eq:propword}, we can compute the entropy of the word coding set $\Omega$ in the context of the complete word string $\mathcal{T}^{\rm wd}$ as
\begin{equation}\label{eq:infoW}
H^{\omega}_0 \equiv H\left({\mathbf p}^{\omega}\right) = - \sum_{\ell=1}^{n^m} p_\ell ^{\omega} \log_2 p_\ell^{\omega}
\end{equation}
where the subscript 0 is used to denote the fact that this compution assumes no autocorrelation among consecutive words.
If consecutive words are autocorrelated, however, then the coding rate must account for the conditional probability $p^{\omega}_{\ell'|\ell}$ of the next word being ${\omega}_{\ell'}$ when the previous word is $\omega_\ell$.  In this case, if $p^{\omega}_{\ell\ell'}$ is the joint probability for the co-occurrence of the word sequence $\omega_\ell$ and $\omega_{\ell'}$ independent of the order in which they occur then the information content is modified to
\begin{equation}\label{eq:infoWAuto1}
H^{\omega}_1  = - \sum_{\ell=1}^{m^n}\sum_{\ell'=1}^{m^n} p^\omega_{\ell\ell'}  \log_2 \frac{p^\omega_{\ell\ell'}}{p^\omega_{\ell}}
\end{equation}
This formulation accounts for only the first level of autocorrelation when higher orders of correlation may occur, as can be assessed by fitting an autoregressive $p$-term  model  AR$(p)$ \cite{biswas2009time} to a word-encoded empirical movement track $\mathcal{T}^{\rm wd}$.

Finally, the set of CAMs has its own entropy which can be computed in terms of the frequency of CAM types $\kappa^\star_c$ in the track $\mathcal{T}^{\kappa^\star}$ when CAM occurrence is not autocorrelated (which, in animal movement ecology, it generally will be). In this case, it follows from Eq~\ref{eq:trackcamassign}  that 
\begin{equation}\label{eq:infoCAM}
     H_0^{\kappa^\star} = - \sum_{c=1}^{k} p^{\kappa^\star}_c \log_2 p^{\kappa^\star}_c \ \ \mbox{and} \ \ H_1^{\kappa^\star} = - \sum_{c=1}^{k} p^{\kappa^\star}_{c c'} \log_2 \frac{p^{\kappa^\star}_{c c'}}{p^{\kappa^\star}_c}
\end{equation}

\section{Information Content and Method Comparisons}\label{subsec:CodeInfo}

The word assignment error $E^{\kappa}$ (Eq~\ref{eq:totError}) and entropy measure $H^{\kappa^\star}$ of a set of rectified CAMs  extracted from a movement track $\mathcal{T}^{\rm loc}$ using the approach illustrated in Fig~\ref{fig:Segmentation}, provides coding scheme measures whereby the performance of two different CAM extraction functors that use different clustering methods and parameter values $\{\mu,n,m,k\}$ (i.e., segment size, StaME number, word size, and CAM number) can be compared.  The coding potential of a set of CAMs, however, also needs to account for the size of the CAMs themselves and not just the entropy $H_0^{\kappa}$ of the CAM set $\mathcal{K}$.

First, we expect that the number of StaMEs groups that can be clearly separated increases with the segment size parameter $\mu$. If we ignore orientation, then for $\mu=2$ only the size (i.e., step length) of StaMEs can vary. If $\mu=3$, then both size and turning angle can vary: apart from size, some StaMEs may have highly acute (almost reversed) turning angles, others closer to right angles, and still others with more obtuse or even rather small turning angles.  The latter is a characteristic of what is called persistent movement.  As $\mu$ increases beyond 3, StaME shapes can become more varied with saw-tooth like elements, circular elements or combinations thereof, and absolute size (i.e., step-size/velocity averaged across all relocation points in the segment) coming into the mix.  The more CAMs that are identified, the greater the potential coding rate since the maximum coding rate of a set of $k$ rectified CAMs occurs when $p^{\kappa^\star}_c = 1/k$, $c=1,\cdots,k$ (Eq~\ref{eq:setOfCAMs}), and is given by 
\begin{equation}\label{eq:maxEntKappa}
    H^{\kappa^\star}_{\max} = - k(1/k)\log_2(1/k) = \log_2 k
\end{equation}

For a track containing $T+1$ relocation points (i.e., including the point 0), however, the number of CAMs of size $m\mu$ that can be used to code this track is $N^{\rm wd}=\lfloor T/(m \mu) \rfloor $ so that the the maximum amount of information that can be coded into track $\mathcal{T}^{\rm CAM}$ is $N^{\rm wd}H_0^{\kappa^\star}$ bits (Table 2). This, of course does not account for the fact that CAM sequences may be autocorrelated, in which case $N^{\rm wd}H_1^{\kappa}$ bits (Eq~\ref{eq:infoCAM}) provides a better estimate of the amount of information that can be coded into a track of length $T$ when segmented into CAMs that each include $m \mu$ steps.

Essentially the task of producing the best (among a comparative group of parameter and clustering method approaches) rectified CAMs from relocation data can be expressed in terms of a family of functors $\mathcal{M}_a$, $a=0,1,2,\cdots,$.  The arguments of these functors are the parameters $\{\mu,n,m,k\}$ and the methods selected to cluster the segments into StaMEs (denoted CM($n$,approach$^\sigma$)) and the words into CAMs (denoted CM($k$,approach$^{\kappa}$)).
Following our earlier notation, our family of functors are mappings of a track $\mathcal{T}^{\rm loc}$ onto a set of  StaMEs $\mathcal{S}$, rectified CAMs $\mathcal{K}^\star$, and error measure $E^{\mathcal{\kappa}}$ all indexed by an approach designator $a$:
\begin{eqnarray}\label{eq:Method}
    \mathcal{M}_a  &: & \mathcal{T}^{\rm loc}_a  \mapsto  \{ \mathcal{T}^{\sigma}_a,\mathcal{T}^{\rm CAM}_a,E^{\mathcal{\kappa}_a}\}, \quad a=1,2,\cdots,  \\
 && \mbox{where } \mathcal{M}_a (\mathcal{T}_a) \equiv \mathcal{M} \big( \mathcal{T}_a;\mu_a, 
\mbox{CM($n_a$,approach$_a^{\sigma}$)},m_a,\mbox{CM($k_a$,approach$^\kappa_a$)} \big) \nonumber
\end{eqnarray}

Beyond using the entropy measures $H^{\kappa^\star_a}$ (Eq~\ref{eq:infoCAM}) and misassignment error rates $E^{\mathcal{\kappa}_a}$ (Eq~\ref{eq:totError}) of different functors $a$ to compare various approaches and identify those with high coding and low error rates, we can compare the CAM distributions ${\bf p}^{\kappa^\star}$ in sets $\mathcal{K}^\star_a$ that have the same number $k$ of CAMs using the information theoretic Kullback-Leibler and Jensen-Shannon divergence measures \cite{manning1999foundations}. The Kullback-Leibler divergence measure allows us to compute the ``distances'' (divergences) between  two distributions.  Specifically, if $\mathbf{p}^1=\left(p^1_1,\cdots,p^1_k \right)$ and $\mathbf{p}^2=\left(p^2_1,\cdots,p^2_k \right)$ are any two discrete distributions with $k$ bins then the Kullback-Leibler measure of the divergence of distribution $\mathbf{p}^2$ from distribution $\mathbf{p}^1$ is
\begin{equation}\label{eq:KLD}
    D_{\rm KL}({\mathbf p}^1:{\mathbf p}^{2})=\sum_{c=1}^{k} p^{1}_c \log_2\left( \frac{p^{1}_c}{p^{2}_c} \right) 
\end{equation}
This measure is not symmetric with respect to the distributions $\mathbf{p}^1=\left(p^1_1,\cdots,p^1_k \right)$ and $\mathbf{p}^2=\left(p^2_1,\cdots,p^2_k \right)$ and is not defined if any of the elements of $\mathbf{p}^2=\left(p^2_1,\cdots,p^2_k \right)$ are zero. The Jensen-Shannon measure solves these issues by first defining the mixture distribution ${\mathbf p}^{\rm mix}$ of ${\mathbf p}^{1}$ and ${\mathbf p}^{2}$ as follows:
if $N_1$ and $N_2$ are the sample sizes of distributions $\mathbf{p}^1$ and $\mathbf{p}^2$ then 
\begin{equation}\label{eq:mixedP}
    {\mathbf p}^{\rm mix}=\left(p^{\rm mix}_1,\cdots,p^{\rm mix}_{k}\right), \ \mbox{ where } \  p^{\rm mix}_c = \frac{N_1 p^{1}_c}{N_1+N_2}  + \frac{N_2 p^2_c}{N_1+N_2}, \ \  c=1,\cdots,k
\end{equation}
In terms of this mixed distribution the Jensen-Shannon divergence is then defined by
\begin{equation}\label{eq:JSD}
    D_{\rm JS}({\mathbf p}^{1},{\mathbf p}^{2})=\frac{1}{2}\Big(D_{\rm KL}({\mathbf p}^{1}:{\mathbf p}^{\rm mix})+ D_{\rm KL}({\mathbf p}^{2}:{\mathbf p}^{\rm mix})\Big) 
\end{equation}
Clearly this measure is now symmetric with respect to distribution $\mathbf{p}^1$ and $\mathbf{p}^2$.  Also, because proportions are nonnegative, $p^{\rm mix}_{c}=0$ if and only if $p^{1}_{c}=0$ and $p^{2}_{c}=0$ the problem of dividing by 0 can be avoided: just remove bin $c$ itself from the comparison.

In the context of the raw CAM set $\mathcal{K}$, we may want to evaluate the difference between any one of the distributions ${\mathbf p}^{\mathcal{W}_c}$ $c=1,\cdots,k $, and the before-cluster distribution ${\mathbf p}^{\omega}$ of words in $\Omega$ by computing $D_{\rm KL}({\mathbf p}^{\mathcal{W}_c}:{\mathbf p}^{\omega})$. More informatively, however, the Jensen-Shannon divergence measure can  be extended to estimate that average divergence across the ensemble of discrete distributions ${\mathbf p}^{\mathcal{W}_c}$ by generating a mixed distribution ${\mathbf p}^{\rm mix}$ using a version of Eq~\ref{eq:mixedP} extended from 2 to $k$ distributions.  Accounting for the relative size $p^{\kappa}_c$ of each of the subsets $\mathcal{W}_c$ (Eq~\ref{eq:setOfCAMs}), the resulting Jensen-Shannon ensemble divergence is then given by
\begin{eqnarray}\label{eq:JSDavg}
D_{\rm JS}^{\rm ens}\big({\mathbf p}^{\mathcal{W}_1},\cdots,{\mathbf p}^{\mathcal{W}_k}\big) &= &\sum_{c=1}^{k} p^{\kappa}_c D_{\rm KL}({\mathbf p}^{\mathcal{W}_c},{\mathbf p}^{\rm mix}) \nonumber \\
    & = &  \sum_{c=1}^{k} p^{\kappa}_c \sum_{\ell=1}^{n^m} p_\ell^{\mathcal{W}_c} \log_2 p_\ell^{\mathcal{W}_c} - \sum_{\ell=1}^{n^m} p_\ell^{\rm mix} \log_2 p_\ell^{\rm mix}  \\
  &=& H^{\omega}\big({\mathbf p}^{\rm mix}\big) - \sum_{c=1}^{k} p^{\kappa}_c H^{\omega}\big({\mathbf p}^{\mathcal{W}_c}\big) \nonumber
\end{eqnarray}
Thus the Jensen-Shannon divergence, used to measure the coding efficacy of track segmentation into raw CAMs, comes down to the difference between the entropy of i) the word distribution before CAM clustering (i.e., $H({\mathbf p}^{\rm mix})$) and ii) a weighted average of the entropies of the $k$ clusters after clustering. The error rate $E^{\mathcal{\kappa}}$ (Eq~\ref{eq:totError}) can then used to assess the error of assigning raw to rectified CAMs.

\section{Discussion}

The formulation provided here only considers information that can be extracted from the relocation time series $\mathcal{T}^{\rm loc}$ itself, whereas these data can be highly context dependent.  The movement of individuals can be both individual-specific and noisy \cite{shaw2020causes}.  It is likely to be affected by the internal state of an individual including hunger \cite{hansen2015effect}, thirst \cite{owen2020movement}, age and sex \cite{luisa2023categorizing}, nutritional state \cite{bazazi2011nutritional}, health \cite{binning2017parasites}, not to mention an individual's ``personality''  or syndromic movement type \cite{spiegel2015going,spiegel2017,hertel2020guide}.  It is also affected by the external state and structure of an individual's immediate \cite{wittemyer2019behavioural} and remembered \cite{fagan2013spatial,polansky2015elucidating,abrahms2019memory} environment, landscape structure and resource distribution \cite{thurfjell2014applications}.  Immediate external factors include current weather, the presence of allies, competitors, and predators, as well as a perception of danger from enemies result in a so-called ``landscape of fear'' \cite{bleicher2017landscape}).   

When auxiliary data such as accelerometer, heart rate, skin temperature and barometric readings are collected from an individual along with GPS location data \cite{nathan2012using,de2018using,seidel2018ecological}, then such data can be incorporated into the formulation to fine tune the informational aspects of movement behavior.  The most obvious way to do this is to condition the occurrence and autocorrelations associated with probabilities and sequencing of StaMEs within words, and CAMs within BAMs and to incorporate these changes into the various informational and divergence measures presented here.

At the BAM level, the motivational state of animal, though influenced by the internal and external variables mentioned above, will also be conditioned by diel-cycle time parameters (e.g., when to sleep, rest, search for food, feed, and so on).  In addition, when an animal commutes on the landscape its direction and speed of heading are likely to be influenced by navigational beacons that relate to an individual's knowledge of its landscape \cite{fagan2013spatial,polansky2015elucidating,abrahms2019memory}.  Such complexities can only be considered once suitable data exist that link an individuals behavior to such beacons, but the next step remains to see how useful the concepts presented here are when applied to real data. These data, of course, must be of sufficient resolution to render the identification of StaMEs a useful enterprise \cite{getz2022simulation,getz2023animal}, rather than the enterprise of directly identifying BAMs using BCPA methods on data that has a resolution of minutes \cite{Teimouri2018clustering} rather than the needed range of seconds or tens of seconds.

The informational approach presented here to segmenting movement tracks of animals, as a way of identifying various factors that influence the structure of animals movement path, provides a rigorous basis for comparing movement tracks of individuals as a function of internal, external and idiosyncratic factors that ultimately determine the movement behavior and trajectory of individuals. The theory and methods articulated provide for the first time a way to rigorously address such questions as ``Does the information content of an individual increase as it learns to navigate its landscape?'' or ``Is the information content of an individual reduced when it is infected with parasites?''

At the cusp of the ``big data'' information age in movement ecology \cite{nathan2022big}, it is only fitting that we have a way of coding movement tracks and assessing and comparing the efficacy of these codes across different parameter sets and clustering algorithms.  \textcolor{black}{Further, as other kinds of data derived from various types of inertial sensors (e.g., gyroscopes, magnetometers, accelerometers) are collected, such data can be used to supplement location data in characterizing segments of fixed size (either time-wise or step-wise).}

\textcolor{black}{Auxiliary data may also become an increasingly important augmentation of relocation time series collected at relatively high frequencies in the context of compensating for observation errors associated with GPS location data.  These location data currently have a CEP (circular error probable) of a few meters, depending on the particular hardware used to collect the data \cite{Teimouri2018clustering}.  Averaging across $\mu$ steps however, will considerably reduce this error \cite{signer2019animal}. The addition of other data types, such as accelerometer data, are likely to increase the statistical robustness of identifying the StaMEs that underpin different types of movement, particular relatively low accelerative (e.g. gliding or striding) versus relatively high accelerative (e.g., jumping or swooping) modes of movement. }

\textcolor{black}{The method presented in this paper hold no matter what type of movement data are used to define a movement track, provided suitable AI clustering methods can be applied to the identification of a robust set of StaMEs. Thus, performance-measure-supported method provides a way forward to implementing a fine-scale structural analyses of animal movement tracks in the decades to come.}

\subsubsection*{Conflict of Interest Statement}

The author declares that the research was conducted in the absence of any commercial or financial relationships that could be construed as a potential conflict of interest.

\subsubsection*{Author Contributions}

N/A

\subsubsection*{Funding}
Funded in part by the A Starker Leopold Endowed Chair in Wildlife Biology at the University of California, Berkeley.

\subsubsection*{Acknowledgments}

I would like to thank Luca Giuggioli and Ran Nathan for inviting me to participate in the programme ‘Mathematics of Movement: an interdisciplinary approach to mutual challenges in animal ecology and cell biology’ held at the Isaac Newton Institute for Mathematical Sciences in 2023 and  where part of the work on this paper was undertaken and supported by the EPSRC Grant Number EP/R014604/1.  

I would like to thank Varun Sethi for constructive discussions regarding the development of the approach taken here and for taking the time to provide comments on this manuscript, as well as Richard Salter and Orr Spiegel for discussions over the past year regarding animal track segmentation.  In addition, and in approximate chronological order, I have benefited over the past 20 years  from the many discussions I have had with the following students and colleagues who have influenced my thinking within the field of spatial and movement ecology: Norman Owen-Smith, Jessica Redfern, Peter Baxter, Chris Wilmers, Paul Cross, Jamie Lloyd-Smith, George Wittemyer, Sadie Ryan, Shirli Bar-David,  David Saltz, Ran Nathan, Craig Tambling, Leo Polansky, Andy Lyons,  Wendy Turner, Steve Bellan, Alex Boettiger, Richard Salter, Werner Kiliam, Royi Zidon, Scott Fortmann-Roe, Miri Tsalyuk,  Briana Abrams, Dana Seidel, Eric Dougherty, Colin Carlson, Krti Tallam, Jason Blackburn, Wayne Linklater, Yen-Hua Huang and Ludovica Luisa Vassat. 

{ \bibliographystyle{unsrtnat}
\bibliography{move}}

\end{document}